\documentclass[conference]{IEEEtran}
\IEEEoverridecommandlockouts

\usepackage{lipsum}
\usepackage{comment}
\usepackage{cite}
\usepackage{amsmath,amssymb,amsfonts}
\usepackage{graphicx}
\usepackage{textcomp}

\usepackage{booktabs}
\usepackage{colortbl}
\usepackage{tabularx}
\usepackage{adjustbox}
\usepackage{multicol}
\usepackage{multirow}

\usepackage{url}

\usepackage{float}
\usepackage{subcaption}

\usepackage{algorithm}
\usepackage{algorithmicx}
\usepackage{algpseudocode}
\usepackage{amsmath} 
\floatname{algorithm}{Algorithm} 

\definecolor{mypink}{rgb}{1,0.9,0.9}
\definecolor{myblue}{rgb}{0.9,0.9,1}
\definecolor{mygreen}{rgb}{0.9,1,0.9}
\definecolor{myyellow}{rgb}{1,1,0.8}

\usepackage{csquotes}

\usepackage{flushend}

\def\BibTeX{{\rm B\kern-.05em{\sc i\kern-.025em b}\kern-.08em
    T\kern-.1667em\lower.7ex\hbox{E}\kern-.125emX}}

\usepackage[shortlabels]{enumitem}
\setlist[itemize]{align=parleft,left=0pt..1em}
\setlist[enumerate]{align=parleft,left=0pt..1.5em}

\definecolor{mypink}{rgb}{1,0.9,0.9}
\definecolor{myblue}{rgb}{0.9,0.9,1}
\definecolor{mygreen}{rgb}{0.9,1,0.9}
\definecolor{myyellow}{rgb}{1,1,0.8}

\usepackage[dvipsnames]{xcolor}
\usepackage{pifont}
\definecolor{tiffany}{HTML}{0abab5}

\usepackage{tcolorbox}
\tcbset{
 top=1pt,
 left=2pt,
 right=2pt,
 bottom=1pt,
 toprule=1pt,
 titlerule=1pt,
 bottomrule=1pt,
 leftrule=1pt,
 rightrule=1pt}

\newcommand{\mycomment}[2]{[\ding{46}~#1:~#2]}

\newcommand{\sinan}[1]{\textcolor{tiffany}{\mycomment{sinan}{#1}}}

\newcommand{\yepang}[1]{\textcolor{blue}{\mycomment{Yepang}{#1}}}

\newcommand{\toolns}{\textsc{VETL}}
\newcommand{\tool}{\textsc{VETL}~}

\begin{document}

\title{Leveraging Large Vision-Language Model For Better Automatic Web GUI Testing}

\author{Siyi Wang$^{1,2,*}$, Sinan Wang$^{1,*}$, Yujia Fan$^2$, Xiaolei Li$^{2,3}$, Yepang Liu$^{1,2,\dagger}$\\

\normalsize $^1$Research Institute of Trustworthy Autonomous Systems, Southern University of Science and Technology, Shenzhen, China\\
\normalsize $^2$Department of Computer Science and Engineering, Southern University of Science and Technology, Shenzhen, China\\
\normalsize $^3$The Hong Kong University of Science and Technology, Hong Kong, China\\
	\normalsize \{12010339,wangsn,12132331,12250067\}@mail.sustech.edu.cn, liuyp1@sustech.edu.cn \\
}

\maketitle

\begingroup\renewcommand\thefootnote{*}
\footnotetext{The two authors contributed equally to this work.}
\endgroup

\begingroup\renewcommand\thefootnote{$\dagger$}
\footnotetext{Yepang Liu is the corresponding author.}
\endgroup

\begin{abstract}
With the rapid development of web technology, more and more software applications have become web-based in the past decades.
To ensure software quality and user experience, various techniques have been proposed to automatically test web applications by interacting with their GUIs.
To achieve high functional coverage, web GUI testing tools often need to generate high-quality text inputs and interact with the associated GUI elements (e.g., click submit buttons).
However, developing a holistic approach that solves both subtasks is challenging because the web GUI context can be complicated and highly dynamic, which is hard to process programmatically.
The recent development of large vision-language models (LVLM) provides new opportunities to handle these longstanding problems.
We in this paper propose \toolns, the first LVLM-driven end-to-end web testing technique.
With LVLM's scene understanding capabilities, \tool can generate valid and meaningful text inputs focusing on the local context, while avoiding the need to extract precise textual attributes.
The selection of associated GUI elements is formulated as a visual question answering problem, allowing LVLM to capture the logical connection between the input box and the relevant element based on visual instructions.
Further, the GUI exploration is guided by a multi-armed bandit module employing a curiosity-oriented strategy.
Experiments show that \tool is effective in exploring web state/action spaces and detecting bugs.
Compared with WebExplor, the state-of-the-art web testing technique, \tool can discover 25\% more unique web actions on benchmark websites.
Moreover, it can expose functional bugs in top-ranking commercial websites, which have been confirmed by the website maintainers.
Our work makes the first attempt of leveraging LVLM in end-to-end GUI testing, demonstrating promising results of this research direction.
\end{abstract}


\section{Introduction}

Modern web technologies have offered users convenient access to various software applications across different platforms~\cite{9401983}.
Ensuring the quality of these web applications is crucial.
For this purpose, automatic web GUI testing (AWGT) methods \cite{fan2023comprehensive} have gained increasing attention.
In an AWGT scenario, a testing agent is programmed for continuous interaction with the website under testing (WUT).
By employing various meta-heuristic strategies, the agent systematically explores the WUT to meet the testing objectives.
For example, WebExplor \cite{zheng2021automatic} formulates web testing as a Markov decision process and solves it with a policy-based reinforcement learning algorithm.
This approach outperforms several other approaches in terms of test adequacy, demonstrating its effectiveness and the potential of intelligent exploration strategies.

One of WebExplor’s limitations is it cannot always generate valid and meaningful inputs for text input areas, while such inputs, like departure/arrival locations for commercial flights in an airline website, are often essential for triggering specific functionalities of the WUT \cite{qtypist}.
In fact, WebExplor can only generate random text inputs, whose quality becomes a major obstacle affecting its test coverage.
QExplore \cite{sherin2023qexplore} explicitly addresses this challenge by introducing a context-aware text generation method.
Following a two-staged framework, this method initially extracts the contextual DOM attributes of the input widget and employs a deep learning model to categorize its general topic.
It then generates appropriate text input based on the topic information.
However, this method requires the HTML elements to have clearly annotated DOM attributes, which are not always present in real-world websites.
To investigate the issue, we examined the top 20 websites as listed by Semrush \cite{semrushWebsitesWorld}.
Specifically, we manually inspected the DOM attributes of their login pages, focusing on those associated with the input boxes for usernames and passwords.
Among the 20 websites, 11 offer the option to login using a phone number.
Surprisingly, only two out of the 11 sites include ``phone'' or ``mobile'' DOM attributes on their input boxes to specify the type of desired input.
QExplore will fail to generate phone numbers as text input on them, which limits its exploration ability.

Recently, the rapid development of large language models (LLMs) have provided new solutions for many challenging software development and maintenance tasks \cite{deng2023large,deng2023large2,qtypist,inputblaster,liu2023make}.
QTypist \cite{qtypist} is the first research work that accomplishes LLM-based text input generation in mobile app GUI testing.
Later, the authors of QTypist proposed InputBlaster \cite{inputblaster}, which aims at producing crash-triggering text inputs via LLM for mobile apps.
These LLM-based research work shows the feasibility of employing pre-trained large models for input generation during testing.
However, the existing techniques cannot be directly applied for effective AWGT due to three reasons: 

\begin{enumerate}

\item \textbf{Web applications typically have richer content than mobile apps, making context extraction challenging.}
Both QTypist and InputBlaster were designated to test mobile apps, which often run on devices with small screens.
They work by extracting local contexts of the input widgets and synthesizing corresponding prompts to LLMs.
Due to different screen sizes and resolutions, web pages generally contain more GUI elements than the corresponding mobile apps \cite{lin2022gui}.
Therefore, extracting necessary and relevant GUI context information is non-trivial for web applications.

\item \textbf{Web GUI elements have less semantic textual attributes than the mobile GUI elements.}
As discussed, even in the top-ranking websites, the DOM attributes annotated on input widgets cannot fully indicate their expected input types.
Moreover, we found that most of these attributes are meaningless, such as \texttt{whsOnd} or \texttt{r-30o5oe}.
Such auto-generated attributes do not reveal the topics of input widgets, but mislead the input topic categorization process and bring cascading inaccuracies to follow-up testing tasks.
Comparatively, mobile GUI attributes, despite often exhibiting absence and significant variability, have proven to serve as informative context for deducing the topics and functionalities of elements\cite{10132285}, \cite{li2020widget}.

\item \textbf{The generated inputs are not fully utilized to guide exploration in these techniques.}
Although these approaches can generate appropriate text inputs based on the given GUI context, they do not effectively utilize these inputs for GUI exploration.
An intuitive example is, the ``search'' button should be clicked immediately after the searching content is typed in, otherwise, the input will be wasted.
Given that large model queries are computationally (and financially) expensive, it is essential for an AWGT approach to explicitly consider how to make the best use of a text input once it is generated.

\end{enumerate}

The recent advancements in multi-modal large language models \cite{yin2023survey} present significant potential to address these gaps \cite{driess2023palm,zhu2023minigpt}.
In particular, large vision-language models (LVLMs) \cite{liu2024visual} bridge the gap between visual and textual information, and have shown exceptional performance in CV tasks, including visual question answering \cite{bazi2023vision}, visual reasoning \cite{achiam2023gpt}, and semantic segmentation \cite{zhang2024vision}.
Observing the impressive capabilities of LVLMs in image understanding, our goal is to explore their application in an AWGT scenario to address the aforementioned challenges.
Our rationale is:
\begin{enumerate}
\item Although web GUI is more complicated than mobile app GUI, given the ability of LVLM on semantic segmentation, we can leverage them to effectively capture and analyze the contextual information needed for visual perception specifically around the targeted input widget.
\item While extracting semantic attributes can be hard, the visual information near the input widget can also indicate its desired content. With LVLM's visual reasoning ability, we can avoid analyzing semantic attributes from the DOM tree, many of which are obfuscated or automatically generated without carrying clear meanings.
\item We formulate the selection of interactive elements as a visual question answering task, and ask an LVLM to determine the most appropriate interacting element based on the context and content of the text input. LVLM's capability of understanding and interpreting textual information enables more contextually relevant interactions within WUT.
\end{enumerate}

In this paper, we propose \toolns, which employs a single LVLM to achieve context-aware text input generation and curiosity-guided GUI exploration for general web applications.
\tool has two major components, the Text Input Generator and the Target Element Selector.
The former extracts contextual information of an input box, captures and annotates a web page's screenshot, and asks an LVLM to generate corresponding text input.
Meanwhile, an interested element will be selected by LVLM, indicating its correlation with the input text box.
Once all input boxes are filled in, the latter component will select the target element to be interacted with, forming an infinite testing loop.
The selection is guided by a multi-armed bandit strategy, aiming to addressing the well-known exploration-exploitation dilemma \cite{wang2020reinforcement}.
Our experimental results show that \tool can effectively explore the state spaces and action spaces of subject websites, outperforming a state-of-the-art AWGT technique WebExplor and three variants of \toolns.
Moreover, \tool is able to find functional bugs in the core functionalities of top-ranking commercial websites, demonstrating its usefulness in web testing tasks.
In summary, this paper makes the following contributions\footnote{Following the open science policy, we release our experiment data in the public repository: \url{https://github.com/sqlab-sustech/VETL-data}.}:
\begin{itemize}
\item To the best of our knowledge, we propose the first LVLM-driven AWGT technique, \toolns, which employs a single LVLM in two AWGT subtasks.
\item Considering the viability and cost of LVLM models in practice, we further offer three \toolns's variants that synergize LLM and heuristics in their core components.
\item We have evaluated \toolns~on four benchmark open-source websites and ten top-ranking commercial websites, which demonstrates the potential of utilizing LVLM in web testing.
\end{itemize}



\section{Preliminaries}

\subsection{Large Language Model}


LLMs are trained on extensive corpora of an ultra-large scale, enabling them to generate diverse forms of text such as articles, conversations, and even programming code \cite{zhao2023survey}.
It can understand \textit{prompts}, which are natural language instructions provided by human users that enforce specific rules and guidelines \cite{liu2023pre,white2023prompt}, and generate relevant and reasonable text that meets the given requirements.

When it comes to the generation of text inputs for GUI testing, the utility of the 175B-parameter GPT-3 \cite{brown2020language} has been demonstrated \cite{qtypist}. 
Since GPT-3 and other members of the GPT family \cite{achiam2023gpt} are closed-source, their remarkable achievements have spurred the emergence of open-source LLMs, such as LLaMA\cite{touvron2023llama}, Alpaca \cite{taori2023stanford} and Vicuna \cite{vicuna2023}.

\subsection{Large Vision-Language Model}

LLMs are designed for \textit{text-only} tasks and lack the ability to handle multimodal information, such as the combination of visual, auditory, or textual inputs. 
To fulfill the demand, significant efforts have been dedicated to develop multimodal LLMs like GPT4 \cite{achiam2023gpt}.
In particular, LVLMs have emerged to tackle compound tasks involving both vision and natural language.
Their examples include LLaVA-1.5 \cite{liu2023llava} and MiniGPT4 \cite{zhu2023minigpt}.

\begin{figure}[t]
\centering
\includegraphics[width=0.698\columnwidth]{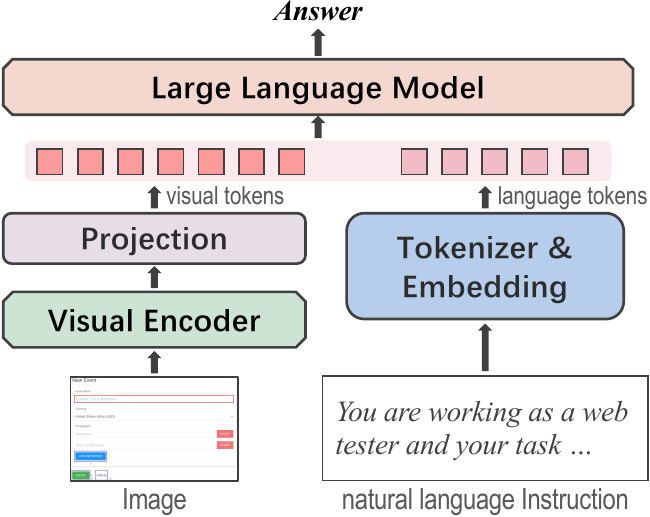}
\caption{Typical LVLM architecture}
\label{fig:LVLM}
\end{figure}

As illustrated in Figure \ref{fig:LVLM}, typical LVLM architecture consists of three components: \textit{visual encoder}, \textit{projection} module, and \textit{LLM}\cite{liu2024survey}.
The visual encoder transforms input image into visual tokens. 
The projection module aligns visual tokens with the word embedding space of the LLM, enabling it to effectively process visual information.
For instance, the LLaVA-1.5 model, known for its impressive multimodal chatting abilities, utilizes CLIP \cite{radford2021learning} as the visual encoder and incorporates  an MLP projection \cite{chen2020improved}.
Its LLM module is a fine-tuned Vicuna-13B \cite{vicuna2023} model, which exhibits superior instruction-following capabilities among public checkpoints.

\section{Approach}


\begin{figure*}[t!]
\centering
\includegraphics[width=\textwidth]{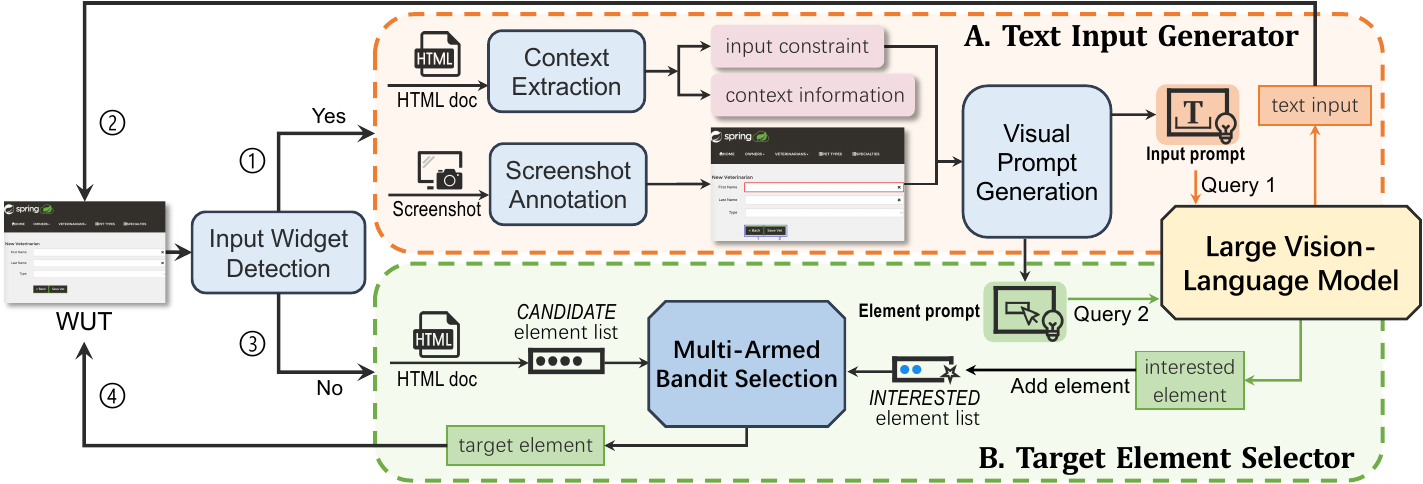}
\caption{Overview of \tool}
\label{fig:webtypist}
\end{figure*}

\subsection{Overview}

In this part, we introduce \toolns, which stands for L\textbf{\underline{V}}LM-empowered w\textbf{\underline{E}}b \textbf{\underline{T}}esting \textbf{\underline{L}}oop.
Its overview is shown in Figure \ref{fig:webtypist}.
\tool has two components, the \textit{Text Input Generator} and the \textit{Target Element Selector}.
They query the same deployed LVLM for their own subtasks, which means this LVLM will not be fine-tuned \cite{deng2023large}. 
We acknowledge that this design choice is influenced by the limitation of our computational resources.
However, once we acquire more powerful GPUs, we will consider assigning different LVLMs to both subtasks and fine-tuning the models to improve their performances. 

In general, \tool continuously interacts with WUT in an infinite loop until the testing objective is achieved or the pre-defined testing resources is exhausted.
For each loop, it observes the current state of WUT by detecting whether it contains text input fields (step \textcircled{\footnotesize{1}} in Figure \ref{fig:webtypist}).
This step can be done by simply searching for semantic DOM elements \cite{w3schoolsHTMLSemantic} in the HTML document.
For the input field detection task, our targets are two kinds of DOM elements, \texttt{input} and \texttt{textarea}.

Once an input box (or widget) is detected, the Input Text Generator will generate a meaningful text input accepted to the that widget (step \textcircled{\footnotesize{2}} in Figure \ref{fig:webtypist}).
Given the widget's containing web page, we extract contextual information from the HTML documentation.
This information includes the widget's global context and local context, as well as the information about the widget itself.
The vision information perceived by the LVLM is provided with a screenshot to the current web page, it teaches the LVLM more about the widget's surrounding environment.
Most importantly, the screenshot will be annotated to better indicate the requirement of the generated text input.

After the input box is filled with the LVLM-generated text, an \textit{interested element} will be selected via a separate LVLM query.
By saying an element is ``interested'', we mean it is pinpointed by LVLM such that the previously generated text can be used by this element.
Such an example can be the ``search'' button after the query input box on a search engine website, or the ``submit'' button at the end of a comment area on a blog post web page.
Ideally, such button should be clicked right after the text input box is filled.
\tool will type in all empty input boxes, since in some scenarios, it requires multiple text inputs to trigger deeper functionalities. 
For example, one should provide both departure and destination locations in a flight booking service.

Once there is no more empty input element presented on the current web page (step \textcircled{\footnotesize{3}} in Figure \ref{fig:webtypist}), \tool will determine which element should be interacted with in this loop.
Given that there can be multiple interested elements corresponding to the previous text input actions, while typing textual input into the text boxes normally does not lead to a web page transition, any of the interested elements can be the target one that will be executed next.
Here, the testing agent should not only consider the actions that can maximize the objective (i.e., the interested elements), but also explore those that may lead to the discovery of previously unknown states.
In the latter case, all interactable elements in the current web page can be the \textit{candidate elements}.
Notice that the candidate elements necessarily contain all interested elements.
The Target Element Selector responds to this task, that is, whether to choose an interested element to maximize the value of previously input texts, or select a candidate element that may potentially expand the exploration space of the current web site.
This situation naturally leads to an \textit{exploration-exploitation dilemma} \cite{wang2020reinforcement}.
We model this problem with a classical \textit{multi-armed bandit} formulation and solve it with a curiosity-guided reward mechanism.
The selected \textit{target element} will then be interacted to continue web exploration (step \textcircled{\footnotesize{4}} in Figure \ref{fig:webtypist}).

\begin{figure}[t!]
\centering

\begin{subfigure}{0.95\linewidth}
\includegraphics[width=0.95\linewidth]{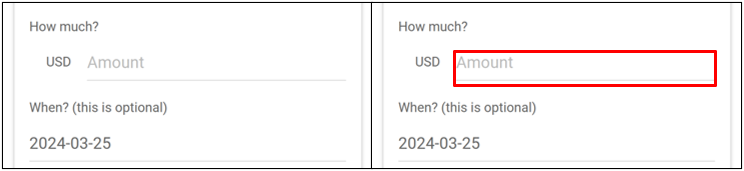}
\caption{screenshot annotation for input prompt}
\label{sf:input}
\end{subfigure}
\par\bigskip
\begin{subfigure}{0.95\linewidth}
\includegraphics[width=\linewidth]{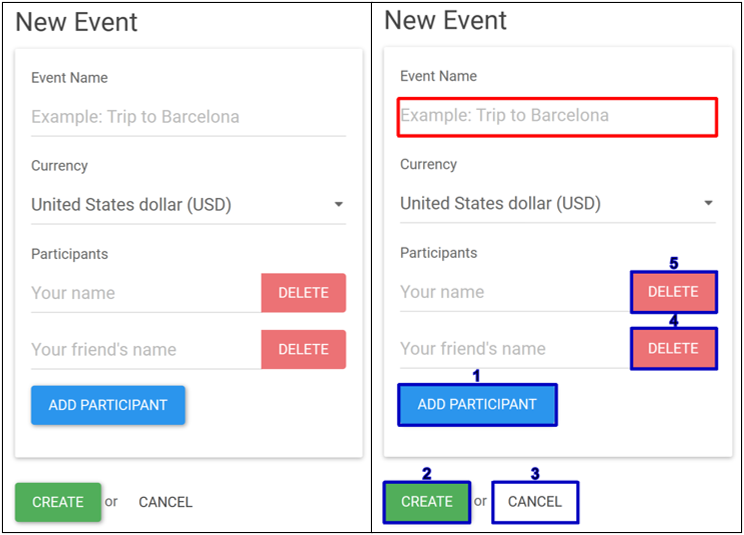}
\caption{screenshot annotation for element prompt}
\label{sf:element}
\end{subfigure}

\caption{Visual prompts on deployed SplittyPie website \cite{githubGitHubTsubiksplittypie} (cropped for presentation purpose)}
\label{fig:visual-prompt}
\end{figure}

Next, we will introduce how \tool generates text inputs and discovers interested elements via LVLM queries, as well as how it chooses the target element from both interested elements and candidate elements.
Figure \ref{fig:visual-prompt} shows the examples of annotated screenshots, which will be fed to LVLM along with the input prompt and element prompt, respectively.
It is worth noting that the screenshots here are manually cropped for presentation purpose, while they will retain their original screen resolutions for LVLM query during testing time.

\subsection{LVLM-Assisted Text Input Generation}

\begin{table*}[t!]
\centering
\caption{\toolns's prompt generation rules (example prompts are derived from Figure \ref{sf:input})}
\label{tab:prompt_template}
\resizebox{\textwidth}{!}{\begin{tabular}{c|ll}
\toprule
\textbf{Template} &
\multicolumn{1}{c|}{\textbf{Input Prompt}} &
\multicolumn{1}{c}{\textbf{Element Prompt}} \\

\midrule

\begin{tabular}[c]{@{}c@{}}Role Playing\\ $\left \langle \textit{RP}\right \rangle $\end{tabular} &
  \multicolumn{1}{l|}{\begin{tabular}[c]{@{}l@{}}
  You are working as a web tester and your task is to generate appropriate text \\
  input for the specified input box on the web page you are browsing.\end{tabular}} &
  \begin{tabular}[c]{@{}l@{}}
  You are working as a web tester and your task is to select a button on a web page \\
  that will be clicked after filling in the input box, such that the button clicking can \\
  submit the filled content and trigger followup web services. \end{tabular} \\

\midrule

\begin{tabular}[c]{@{}c@{}}Visual Information\\ $\left \langle \textit{VI}\right \rangle $\end{tabular} &
  \multicolumn{1}{l|}{\begin{tabular}[c]{@{}l@{}}
  You are provided with a screenshot of the web page you are browsing, the  \\
  input box to be filled is marked with a \textcolor{red}{red} frame. \end{tabular}} &
  \begin{tabular}[c]{@{}l@{}}
  You are provided with a screenshot of the web page you are browsing, the input \\ 
  box to be filled is marked with a \textcolor{red}{red} frame and the buttons to be selected are\\ 
  marked with \{\textit{Range of Button Numbers}\} and \textcolor{blue}{blue} frames. \end{tabular} \\

\midrule

 &
  \multicolumn{2}{c}{You are viewing the web page entitled: \{\textit{HTML Title}\}.} \\ 
\multirow{-2}{*}{\begin{tabular}[c]{@{}c@{}}Global Context\\ $\left \langle \textit{GC}\right \rangle $\end{tabular}} &
  \multicolumn{2}{c}{{\color[HTML]{A6A6A6} e.g. You are viewing the web page entitled: \textit{SplittyPie}.}} \\

\midrule

 &
  \multicolumn{2}{c}{This text is placed near the input box: \{\textit{Text of Nearby Element}\}.} \\ 
\multirow{-2}{*}{\begin{tabular}[c]{@{}c@{}}Local Context\\ $\left \langle LC\right \rangle $\end{tabular}} &
  \multicolumn{2}{c}{{\color[HTML]{A6A6A6} e.g. This text is placed near the input box: \textit{How much? USD}.}} \\

\midrule

 &
  \multicolumn{1}{l|}{1. The input has type \{\texttt{type} \textit{Attribute of Input Box}\}.} &
   \\
 &
  \multicolumn{1}{l|}{{\color[HTML]{A6A6A6} e.g. The input has type \textit{number}.}} &
   \\
 &
  \multicolumn{1}{l|}{2. The input is about \{\texttt{id}\ /\ \texttt{placeholder}\ /\ \texttt{name} \textit{Attribute of Input Box}\}.} &
   \\
 &
  \multicolumn{1}{l|}{{\color[HTML]{A6A6A6} e.g. The input is about \textit{amount}.}} &
  \multirow{-1}{*}{The input box will be filled with: \{\textit{Generated Text Input}\}.} \\
 &
  \multicolumn{1}{l|}{3. The input can be \{\texttt{value} \textit{Attribute of Input Box}\}.} &
  {\color[HTML]{A6A6A6} } \\
 &
  \multicolumn{1}{l|}{{\color[HTML]{A6A6A6} /}} &
  {\color[HTML]{A6A6A6} } \\
 &
  \multicolumn{1}{l|}{4. The input has the following constraints: \{\textit{Inferred Constraints of Input Box}\}.} &
  {\color[HTML]{A6A6A6} } \\
\multirow{-8}{*}{\begin{tabular}[c]{@{}c@{}}Input Widget\\ $\left \langle \textit{IW}\right \rangle $\end{tabular}} &
  \multicolumn{1}{l|}{{\color[HTML]{A6A6A6} e.g. The input has the following constraints: \textit{minimum value is 0}.}} &
  \multirow{-7}{*}{{\color[HTML]{A6A6A6} e.g. The input box will be filled with: \textit{199}.}} \\

\midrule

\begin{tabular}[c]{@{}c@{}}Output Structure\\ $\left \langle \textit{OS}\right \rangle $\end{tabular} &
  \multicolumn{1}{l|}{\begin{tabular}[c]{@{}l@{}}
  Your job is to generate appropriate text for the highlighted input box. You \\ 
  should only return the generated text without any explanation. Use the \\ 
  following format for your answer: Generated Input Text: \textbf{{[}answer{]}}\end{tabular}} &
  \begin{tabular}[c]{@{}l@{}}
  Your job is to select one of the labeled buttons and return the number on it \\ 
  without any explanation. Use the following format for your answer: Selected \\
  Button Number: \textbf{{[}answer{]}}\end{tabular} \\

\midrule\midrule

\textbf{\begin{tabular}[c]{@{}c@{}}Rules of Prompt\\ Generation\end{tabular}} &
  \multicolumn{1}{c|}{\cellcolor{orange!20}$\left \langle \textit{RP}\right \rangle $+$\left \langle \textit{VI}\right \rangle $+$\left \langle GC\right \rangle $+$\left \langle LC\right \rangle [$+$\left \langle \textit{IW}\right \rangle]_n $+$\left \langle \textit{OS}\right \rangle $} &
  \multicolumn{1}{c}{\cellcolor{LimeGreen!20}$\left \langle \textit{RP}\right \rangle $+$\left \langle \textit{VI}\right \rangle $+$\left \langle GC\right \rangle $+$\left \langle LC\right \rangle $+$\left \langle \textit{IW}\right \rangle $+$\left \langle \textit{OS}\right \rangle $} \\

\bottomrule
\end{tabular}
}
\end{table*}

\subsubsection{Annotating Screenshot for Visual Prompting}

For vision models, visual prompting \cite{zou2024segment} allows users marking labels on the input images for accelerating downstream tasks.
In \toolns's use case scenario, we use visual prompting method to highlight the target input box in a single LVLM query.
Taking Figure \ref{sf:input} as an example.
While there are multiple input boxes appear on the screenshot, we mark ``Amount'' with a red frame to tell LVLM generate text input for that specific field.
LVLM can also learn this indication through the prompting message, as shown by the template $\textit{VI}$ in Table \ref{tab:prompt_template}.

Each time, \tool generates one text input for a specific input box, thus only a single color will be used for labeling screenshot.
It is also possible to generate text inputs for multiple input boxes, such as generating inputs for both ``Amount'' and date field in one LVLM query (Figure \ref{sf:input}).
In such cases, it becomes necessary to label the input boxes with different colors (or assign numbers to boxes) and establish mappings between the input boxes and their corresponding colors (or numbers), which should be represented in the prompting message.
However, based on our manual testing, this approach was found to be less effective as the LVLM struggles to accurately distinguish between these different input boxes.
Nevertheless, we believe that as larger models continue to advance, this situation can be resolved.
Therefore, we will revisit this alternative approach in the future.

\subsubsection{Text Prompt Template}

In this work, \textit{input prompt} refers to the prompt message sent to LVLM for text input generation.
As indicated by \cite{qtypist}, the input box's contextual information helps the language model better understand input semantics on the current GUI.
\tool also employ a context-aware prompt generation rule, as described in Table \ref{tab:prompt_template}.

An input prompt starts with a role playing preamble \cite{shanahan2023role}, follows by a visual information description illustrating the accompanied image.
Similar to \textsc{QTypist}'s approach, we consider two types of contextual information:
the global context about the WUT's metadata, and the local context about nearby message of the input box.
Here, we extract the WUT's title as global context, as it often provides a concise summary or description of the web page content.
Local context information refers to textual data nearest to the input box, which aids in comprehending the input semantics.
We adopt a simple heuristic rule to extract the local context.
It will select the closest neighboring node in the DOM tree to the input box that contains non-empty inner HTML text.
Next is description of the input box itself, which will be depicted in Section \ref{sssec:constrain}.
Lastly the prompt specifies the requirements for the output format.
A typical input prompt query will be like (together with Figure \ref{sf:input} as the visual input):

\begin{displayquote}
\footnotesize

\textbf{\small User Prompt}: You are working as a web tester and your task is to generate appropriate text input for the specified input box on the web page you are browsing. You are provided with a screenshot of the web page you are browsing, the input box to be filled is marked with a red frame. You are viewing the web page entitled: SplittyPie. This text is placed near the input box: How much? USD. The input has type number. The input is about amount. The input has the following constraints: minimum value is 0. Your job is to generate appropriate text for the highlighted input box. You should only return the generated text without any explanation. Use the following format for your answer: Generated Input Text: [answer]

\textbf{\small LVLM Answer}: Generated Input Text: 199

\end{displayquote}

\subsubsection{Extracting Input Box Information and Constraints}
\label{sssec:constrain}

\begin{table}[t!]
\centering
\caption{Rules of extracting input box constraints}
\label{tab:constraint_pattern}
\resizebox{\columnwidth}{!}{
\begin{tabular}{m{4.2em}<{\centering}|m{3.5em}<{\centering}|m{4em}<{\centering}|m{18.5em}}
\hline
\textbf{DOM}\newline\textbf{Element} & \textbf{Type}\newline\textbf{Attribute} & \textbf{Constraint Attribute} & \multicolumn{1}{c}{\textbf{Constraint Pattern}} \\
\hline
\multirow{7}[21]{*}{
\textless input\textgreater} & \multirow{2}{*}{text} & maxlength & maximum length of text is \{\textit{maxlength}\} \\
\cline{3-4}    \multicolumn{1}{c|}{} & \multicolumn{1}{c|}{}    & minlength & minimum length of text is \{\textit{minlength}\} \\
\cline{2-4}    \multicolumn{1}{c|}{} & \multirow{2}{*}{password} & maxlength & maximum length of password is \{\textit{maxlength}\} \\
\cline{3-4}    \multicolumn{1}{c|}{} & \multicolumn{1}{c|}{} & minlength & minimum length of password is \{\textit{minlength}\} \\
\cline{2-4}    \multicolumn{1}{c|}{} & email & multiple & multiple emails are allowed, with each email separated by a comma \\
\cline{2-4}    \multicolumn{1}{c|}{} & \multirow{3}{*}{number} & max   & maximum value of number is \{\textit{max}\} \\
\cline{3-4}    \multicolumn{1}{c|}{} & \multicolumn{1}{c|}{} & min   & minimum value of number is \{\textit{min}\} \\
\cline{3-4}    \multicolumn{1}{c|}{} & \multicolumn{1}{c|}{} & step  & number interval is \{\textit{step}\} since \{\textit{min}\} \\
\cline{2-4}    \multicolumn{1}{c|}{} & search & \slash{} & \slash{} \\
\cline{2-4}    \multicolumn{1}{c|}{} & tel   & pattern & telephone number has regular expression pattern \{\textit{pattern}\} \\
\cline{2-4}    \multicolumn{1}{c|}{} & url   & \slash{} & \slash{} \\
\hline
\textless textarea\textgreater & \slash{} & \slash{} & multi-line input is allowed \\
\hline
\end{tabular}

}
\end{table}

For an input box, \tool synthesizes its corresponding prompt according to $\textit{IW}$ template in Table \ref{tab:prompt_template}.
It searches for five DOM attributes, \texttt{type}, \texttt{id}, \texttt{placeholder}, \texttt{name}, and \texttt{value}, and generates their descriptions (if presented).
These attributes are commonly associated with \texttt{<input>} HTML elements, while \texttt{<textarea>} elements will also hold some of these attributes to specify their metadata.
For example, in Figure \ref{sf:input}, the labeled input box has a type of ``number'', and an attribute \texttt{placeholder} with value ``Amount'', hence the corresponding $\textit{IW}$ prompt will be synthesized with the first and the second rules.
Since it does not have a \texttt{value} attribute, the third rules will not be applied.

To ensure the legality of the generated texts, \tool will further generate textual descriptions for the input boxes' constraints.
We only consider those constraints specified in HTML attributes.
According to the HTML standard \cite{mozillainputInput}, we derived a set of constraint patterns and summarize them in Table \ref{tab:constraint_pattern}.
In Figure \ref{sf:input}, the labeled input box has a type of ``number'', and a \texttt{min} attribute with value ``\texttt{0}''.
This leads to a constraint description: ``minimum value is 0''.
The other constraints can also be obtained accordingly.

\subsection{LVLM-Assisted Interested Element Selection}

\subsubsection{Screenshot Annotation}

We use \textit{element prompt} to ask LVLM to select an interested element for the given input box and the corresponding generated text.
Similar to the input prompt, we also annotate on the screenshot.
However, this time, our target is to select an interactive element, which can utilize the generated text input and trigger successor functionalities.
Therefore, we labeled all clickable elements on the screen and ask LVLM to select one of them.
Figure \ref{sf:element} shows such an example, in which all \texttt{<button>} elements are labeled with blue frames and their corresponding button numbers.
We retain the red frame on the selected input box so that it visually distinguishes itself from other input boxes.

\subsubsection{Prompt Generation}

Table \ref{tab:prompt_template} also shows the template for element prompt.
It has a similar structure to the input prompt, which contains role playing, visual information about the input image, global and local contexts, information about the input box, and a specification of output format.
A major difference is that, the input widget information is replaced by the generated text, as the DOM attributes are less important once we obtained the text input.
Similarly, an example query of element prompt will be like (with Figure \ref{sf:element}):

\begin{displayquote}
\footnotesize

\textbf{\small User Prompt}: You are working as a web tester and your task is to select a button on a web page that will be clicked after filling in the input box, such that the button clicking can submit the filled content and trigger followup web services. You are provided with a screenshot of the web page you are browsing, the input box to be filled is marked with a red frame and the buttons to be selected are marked with 1,2,3,4,5 and blue frames. You are viewing the web page entitled: SplittyPie. This text is placed near the input box: How much? USD. The input box will be filled with: 199. Your job is to select one of the labeled buttons and return the number on it without any explanation. Use the following format for your answer: Selected Button Number: [answer]

\textbf{\small LVLM Answer}: Selected Button Number: 2

\end{displayquote}

\subsection{MAB For Target Element Selection}

At this stage, \tool has filled in all empty input boxes on the current web page.
Meanwhile, by querying for LVLM per text input, a list of interested elements is selected by LVLM for handling the input texts.
However, in an AWFT scenario, choosing the proper \textit{target element} to be executed in the next step can be challenging for two reasons:
first, the interested elements can lead to different functional points, while only one element can be executed in a single-browser environment;
second, those non-interested elements can also extend the exploration space and improve the testing coverage, thus they should also be considered.
Taking Figure \ref{sf:element} as the example.
Assume that the ``create'' button is selected as the interested element with the ``event name'' input box, and the ``delete'' button is interested to a ``participants'' input box.
Both interested elements can experience different use case scenarios of the ``new event'' functionality.
They shall be selected to be the target one with different weights, as creating new data entries is prefer for most cases.
On the other hand, although the ``add participant'' button is not selected by LVLM to be interested, we still want it to have a certain probability of being chosen, ensuring the execution of this particular functional point.
As such, we call all the available elements in the current web page as \textit{candidate elements}.
The Target Element Selector will choose one to be the target element from the candidate element set, while paying special attention to its subset, i.e., the interested elements.
This decision making process faces an exploration-exploitation dilemma \cite{sutton2018reinforcement}.
In other words, we should decide choosing an action (i.e., the target element) that may either
1) \textbf{explore} the WUT with no guidance to potentially discover unknown web states, or
2) \textbf{exploit} the profit-gaining action greedily as advised by LVLM previously.

We model this problem with a multi-armed bandit (MAB) formulation.
The MAB model has been extensively used in automated testing research work to tackle decision making tasks, such as test case prioritization \cite{lima2020multi} or fuzzy mutator scheduling \cite{deng2023large}.
During the testing session, \tool will continuously interact with all elements $E$ on the WUT, which can be regarded as $|E|$ arms.
Every time, an arm playing action yields a \textit{reward} indicating its exploration capability, where the objective is to maximize the accumulated reward upon the pre-defined testing budget exhausted.
As the expected reward for each element is unknown before testing, it will be learnt iteratively as a function (a.k.a., Q-table) following an $\varepsilon$-greedy strategy.
Algorithm \ref{alg:mab} shows this decision making and learning process in a single iteration.

\begin{algorithm}[t!]
\caption{Multi-Armed Bandit Target Element Selection}
\label{alg:mab}

\begin{algorithmic}[1]
\Require Q-table $Q: E\to\mathbb{R}$,

\Statex     element counting table $count: E\to\mathbb{N}$,

\Statex     set of interested elements $\textit{E}_\textit{I}$ ($\textit{E}_\textit{I}\subseteq \textit{E}$),

\Statex     set of candidate elements $\textit{E}_\textit{C}$ ($\textit{E}_\textit{C}\subseteq \textit{E}$),

\Statex     exploration probability $\varepsilon$

\If{$\texttt{random}() < \varepsilon$}              \label{line:random}
    \State $\textit{target} \gets e\in \textit{E}_\textit{C}$ with uniform probability              \label{line:explore}
\Else
    \State $\textit{target} \gets e\in \textit{E}_\textit{I}$ with $\Pr[e]=\dfrac{\exp{(Q(e))}}{\sum_{e'\in \textit{E}_\textit{I}} \exp{(Q(e'))}}$              \label{line:exploit}
\EndIf

\State click on element $\textit{target}$               \label{line:click}
\State $r \gets$ number of new elements in current web page             \label{line:reward}
\vspace{1ex}
\State $Q(\textit{target}) \gets \dfrac{Q(\textit{target})\times count(\textit{target})+r}{count(\textit{target})+1}$               \label{line:update_q}
\vspace{1ex}
\State $count(\textit{target}) \gets count(\textit{target})+1$              \label{line:update_c}
\end{algorithmic}

\end{algorithm}

The algorithm takes five input parameters: the Q-table $Q$ estimating the expected reward of applying particular element, and the counting table $count$ recording the times of its execution; the sets of interested elements $\textit{E}_\textit{I}$ and candidate elements $\textit{E}_\textit{C}$, and an exploration probability $\varepsilon$.
Each time, \tool take an explore action with a probability of $\varepsilon$ (line \ref{line:random}), which will randomly select an element from all candidate ones (line \ref{line:explore}).
In contrast, it will also exploit the past experience, i.e., the Q-table (line \ref{line:exploit}).
Instead of choosing a specific interested element, \toolns's strategy is to perform a weighted sampling over $\textit{E}_\textit{I}$, such that those having higher Q-values is more likely to be selected.
The probability of taking a particular interested element is calculated through Softmax function \cite{tokic2011value}, which can be viewed as a simplified version of the Gumbel-Softmax method adopted by \textit{WebExplor} \cite{zheng2021automatic}.
After applying the selected target element (line \ref{line:click}), we calculate a reward for the element.
In an AWGT task, the ultimate goal is to completely visit the whole exploration space representing the WUT.
This means a ``good'' action should discover unknown web states as much as possible.
We quantify this extent with the number of previously unknown elements (line \ref{line:reward}) in the resulting web page, which is also inspired by the curiosity-based rewarding mechanism of \textit{WebExplor}.
With the target element's reward, we can update its Q-value via incremental formula (line \ref{line:update_q}) and increase its apply count by 1 (line \ref{line:update_c}).

Both $Q$ and $count$ are initialized as functions with zero values, and their function values will be updated during testing.
Element records will only be retained within a testing session to ensure the effective guidance of the curiosity mechanism.

\subsection{Fallback Strategies}

While \toolns's main loop is designed to run indefinitely, there might be certain situations that could disrupt its normal procedure.
Such situations arise when 1) LVLM's output does not conform to our instructed formats, such as missing the leading phrase ``Generated Input Text:'', or returning a non-existing button number, and 2) there are no input boxes on the current web page.
For the first case, we will directly use the output text for the text input generation subtask, regardless of its content.
Or we will simply omit the selection of interested element for this round.
It is worth mentioning that such cases are rare during our experiment.
In the second case, \tool will simply take the explore action, that is, randomly select an element from the candidate element list as the target one.

\section{\toolns's Variants}

In real-world web development scenarios, the usage of LVLM can be costly.
To address this issue, we introduce three variants of \toolns.
They differ in their selection of methods utilized for the two major subtasks: text input generation and interested element identification.

\subsection{$\textit{VETL}_\textit{V1}$: Single LVLM Query}

The first variant of \tool replaces LVLM-based interested element selection by a nearest neighboring element selector.
According to the principle of spatial locality \cite{li2022cross}, the most relevant interactive element to an input box is likely to appear in its nearby region.
Following this principle, this variant considers the button closest to the input box as its interested element.
As a result, the variant, which we named $\textit{VETL}_\textit{V1}$, will perform only one LVLM query for text input generation in each test loop.
Meanwhile, the interested element will be selected over the nearby buttons $B$ based on their distances in a DOM tree structure \cite{zhou2021simplified}:
$$\textit{interested}\gets\mathop{\arg\!\min}_{e\in B}\texttt{distance}(e,\textit{input\_box})$$
After that, the target element will still be selected according to the original MAB-based procedure (Algorithm \ref{alg:mab}).

\subsection{$\textit{VETL}_\textit{LV}$: Combined LLM-LVLM Queries}

The second variant generates text inputs via LLM instead of LVLM.
The idea is to replicate {QTypist}'s LLM-based input generation for replacing \toolns's Text Input Generator.
Since \toolns's input prompt template follows similar structure with {QTypist}'s linguistic patterns, we reuse the input prompt while replace the LVLM as LLM:
$$\left \langle \textit{RP}\right \rangle + \left \langle GC\right \rangle + \left \langle LC\right \rangle [+ \left \langle \textit{IW}\right \rangle]_n + \left \langle \textit{OS}\right \rangle$$
Notice that we remove the visual information section $\textit{VI}$ from the prompt because LLM does not receive image input.

Besides the LLM-based text input generation module, the interested element will still be selected by LVLM with element prompt.
Therefore, we call this variant $\textit{VETL}_\textit{LV}$, as it performs one LLM query for text input generation and one LVLM query for interested element selection.

\subsection{$\textit{VETL}_{L}$: LLM-only Query}

The third variant combines the replacement strategies utilized in the previous two variants.
Specifically, given an input box, it will generate its text input by LLM, and will select an interested element which is its nearest neighboring button.
This variant is called $\textit{VETL}_{L}$, since it only queries LLM once for each test loop.

\section{Implementation}

We implemented \tool with Selenium \cite{seleniumSelenium} for basic web elements discovery and UI automation.
The screenshot is captured and annotated using the Selenium-shutterbug library \cite{githubGitHubAssertthatseleniumshutterbug}.
\toolns's underlying VLM is LLaVA-1.5 \cite{liu2023llava}, which achieves state-of-the-art on common CV tasks.
Due to limited computational resources, we deployed and utilized only the 7B model.
To fairly compare the performance between VLM and LLM, we used Vicuna \cite{vicuna2023} as the LLM model for the two variants $\textit{VETL}_\textit{LV}$ and $\textit{VETL}_\textit{L}$, because Vicuna is the base model of LLaVA-1.5.
These large models were deployed on a bare-metal server with two NVIDIA TITAN Xp GPUs.
We set $\varepsilon=0.3$ because it performed well in our pilot runs.

\section{Evaluation}

\subsection{Research Questions}

To evaluate \toolns, we conduct a set of experiments to investigate the following three research questions:

\begin{enumerate}
\item \textbf{RQ1 (Exploration Capability)}: Can \tool outperform its three variants and the state-of-the-art approach in automatic web GUI exploration?
\item \textbf{RQ2 (Ablation Study)}: How does \tool perform when its core components are removed or altered?
\item \textbf{RQ3 (Tool Usefulness)}: Does \tool demonstrate practical usefulness in real-world web testing tasks?
\end{enumerate}

\subsection{Experimental Subjects}

Our experimental subjects include two types of WUTs:
\begin{itemize}
\item \textbf{Open-source benchmark websites}:
We downloaded the source code of four websites and deployed them on our servers. These websites are: DimeShift \cite{githubGitHubJekakiselyovdimeshift}, PetClinic \cite{githubGitHubSpringprojectsspringpetclinic}, Phoenix Trello \cite{githubGitHubBigardonephoenixtrello}, and SplittyPie \cite{githubGitHubTsubiksplittypie}.
They have been used as subject websites by many existing research work \cite{zheng2021automatic,biagiola2019diversity,fan2023comprehensive,Stocco2016APOGENAP} on automatic web testing.
\item \textbf{Top-ranking commercial websites}:
We selected ten top-ranking websites \cite{dataforseo1000Websites} to evaluate \tool on real-world web applications.
Specifically, we selected such a commercial website if it:
1) can be stably visited,
2) has at least three different input boxes,
and 3) does not pop-up anti-robot systems.
The full list of these websites can be found on our data publicity repository.
\end{itemize}

\subsection{RQ1: Exploration Capability}

\begin{table*}[htp]
\centering
\caption{Comparison of \toolns, \textit{WebExplor}, and \toolns's variants}
\label{table:rq1}
\resizebox{0.9\textwidth}{!}{
\renewcommand{\arraystretch}{1.2}
 \begin{tabular}{|c|c|c|c|c|c|c|c|c|c|c|}
\hline
\multirow{2}*{WUT} & \multicolumn{5}{c|}{\#Visited Web State} & \multicolumn{5}{c|}{\#Discovered Web Action} \\
\cline{2-11}          & VETL  & \textit{WebExplor} & $ \textit{VETL}_\textit{V1}$  & $ \textit{VETL}_\textit{LV}$ & $ \textit{VETL}_{L}$ & VETL  & \textit{WebExplor} & $ \textit{VETL}_\textit{V1}$ & $ \textit{VETL}_\textit{LV}$ & $ \textit{VETL}_{L}$ \\
\hline
DimeShift & \textbf{10.8} & 3.6   & 7.2   & 7.8   & 8.6   & \textbf{121.8} & 96.2  & 49.2    & 45    & 71 \\
PetClinic & \textbf{9} & 3.8   & 8.2   & 8     & 8.4   & \textbf{27.4} & 16.2  & 24    & 25.8  & 23.4 \\
Phoenix & \textbf{7} & 5.2   & 4.8   & 6     & 6.4   & \textbf{35} & 33.6  & 30.8  & 31.2  & 32.2 \\
SplittyPie & \textbf{5.6} & 2.2   & 3     & 3     & 2.4   & \textbf{24} & 20.6  & 15.6  & 17.8  & 14.4 \\
\hline\hline
avg. & \textbf{8.1}	& 3.7	& 5.8	& 6.2 &	6.45 &	\textbf{52.05} &	41.65 &	29.9 &	29.95 &	35.25\\
\hline

\end{tabular}

}
\end{table*}

\subsubsection{Setup}

Following previous research work \cite{mesbah2012crawling}, we use \textit{the numbers of web states and web actions} to measure the exploration capability of an AWGT approach.
Intuitively, within the given testing budget, a better GUI explorer shall visit more unique web states and discover more executable web actions.
To unify the measurement of metrics for different techniques, we use the numbers of opened URLs as the numbers of visited web states, and calculate the sum of all interactive GUI elements as the numbers of discovered web actions.

Besides \toolns's three variants, we also compare it with WebExplor \cite{zheng2021automatic}, the state-of-the-art AWGT approach.
Since its source code and software is unavailable, we implemented our version of WebExplor according to its paper.

Large model queries are time-consuming.
For this reason, we do not restrict a fixed testing time.
Instead, we set a total number of 200 web actions (including inputs and button clicks) for each single run.
This cost 20.5 minutes on average for each run of \tool in our experiment.
All experiments are repeated five times to reduce randomness.

\subsubsection{Result}

Table \ref{table:rq1} presents the results of running the five approaches on four open-source benchmark websites.
It shows that \tool consistently outperforms the others in terms of both visited states and discovered actions.

Considering the visited states, WebExplor always exhibits inferior performance compared to the other methods.
Since these WUTs are all content management systems, they will usually redirect to new states (web links) after submitting forms.
Comparing to \tool and its variants, WebExplor only generates random texts and does not effectively utilize the input texts.
As a result, it struggles to successfully submit the filled forms, limiting its exploration capability.


As for the discovered actions, comparing to WebExplor, \tool uncovers 25\% ($=(52.05-41.65)/41.65$) more unique actions, outperforming it by a significant margin.
However, WebExplor performs better than the three variants on three WUTs.
When the WUT is not fully explored, driven by curiosity, WebExplor tends to visit states containing more web actions, even if it cannot complete all input boxes.
While the other three WUTs initially have more interactive elements, on PetClinic, however, the number of interactive elements gradually increases as more data entries being created.
WebExplor's inability of accurately filling out and submitting forms hinders this process.
Consequently, it performs poorly on PetClinic.


\begin{figure}[t!]
\centering
\includegraphics[width=0.95\columnwidth]{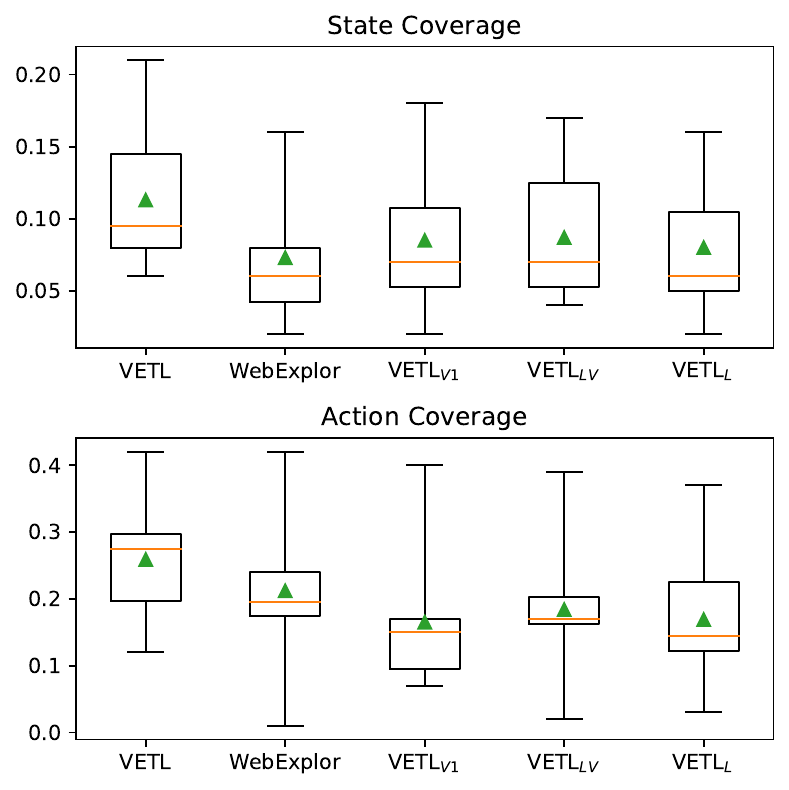}
\caption{Performance comparison on commercial websites}
\label{fig:rq1-commercial}
\end{figure}

Figure \ref{fig:rq1-commercial} illustrates the results on 10 commercial websites.
The x-axis represents the compared approaches, and the y-axis corresponds to their achieved state coverage and action coverage.
Since the total number of states of a commercial website is unknown to us, we estimate it by summing up all discovered states in our experiments.
Then, we calculate the estimated state coverage by taking the ratio of discovered states to the estimated total states.
The same calculation method is also applied to estimate the action coverage.

Based on Figure \ref{fig:rq1-commercial}, it can be observed that \tool achieves the best performance on the commercial websites.
In terms of state coverage, \tool and its three variants outperform WebExplor, with \tool achieving a mean coverage of 0.11, which is 57\% ($=(0.11-0.07)/0.07$) higher than WebExplor's mean coverage of 0.07.
This can be attributed to a similar reason as observed on open-source WUTs: the subject WUTs contain a significant number of input fields that can be filled and submitted to access new web states.
As for action coverage, \tool achieves a mean of 0.26, surpassing WebExplor's coverage of 0.21.
However, WebExplor remains notable and ranks just below \toolns, as its curiosity-driven strategy enables it to explore a significant number of web actions.

According to the statistics from Figure \ref{fig:rq1-commercial}, \tool performs better than $\textit{VETL}_\textit{LV}$, and $\textit{VETL}_\textit{V1}$ shows slightly better results than $\textit{VETL}_\textit{L}$ in terms of the mean, median, maximum, and minimum values of both metrics.
Due to the simpler input scenarios and the availability of richer semantic information within DOM elements on open-source WUTs, the generated prompts effectively guide LLM to generate high-quality inputs.
On the contrary, the vision-assistance approaches demonstrate their superiority on large-scale websites. 
Consequently, \tool and $\textit{VETL}_\textit{V1}$ exhibit more significant advantages on commercial websites than small-sized open-source websites.

\begin{tcolorbox}
\textit{\textbf{Answer 1}}:
\tool outperforms the state-of-the-art technique by exploring 25\% more web actions in benchmark websites.
Its vision components enhance scene understanding, leading to a better performance than LLM-based approaches.
\end{tcolorbox}

\subsection{RQ2: Ablation Study}

\subsubsection{Setup}

In this RQ, we compare \tool against three ablation groups:
a) the Text Input Generator ablation group, where it is individually replaced by a random input generator,
b) the Target Element Selector ablation group, where it is replaced by random element selector,
and
c) the combined ablation group, where the Text Input Generator and the Target Element Selector are simultaneously altered.
The objective is to evaluate the independent contributions of both components within \tool, as well as their collective effects on the overall exploration performance of \tool in web testing.

Similar to RQ1, we evaluate the effectiveness of ablation groups by measuring the numbers of visited web states and discovered web actions.
Each individual run is constrained to a total of 200 actions and is repeated five times.

\subsubsection{Result}


\begin{figure}[t!]
\centering
\includegraphics[width=0.985\columnwidth]{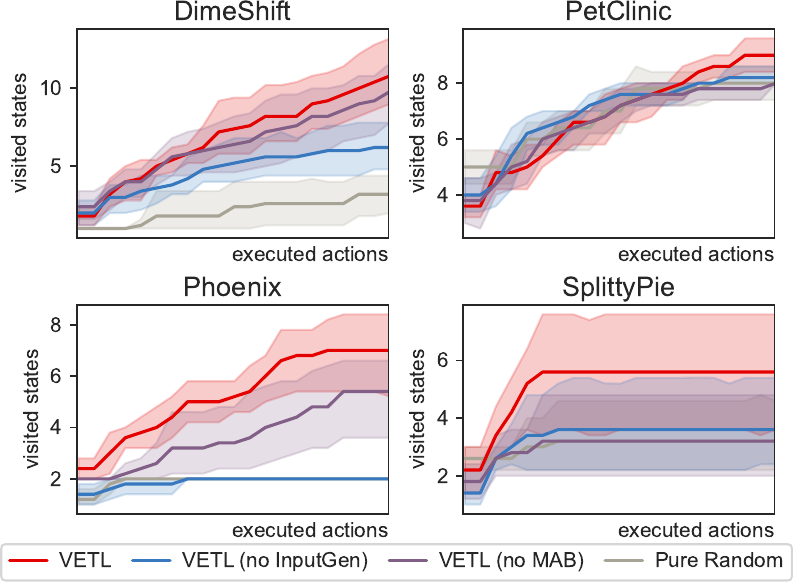}
\caption{Comparison between \tool and ablation groups}
\label{fig:rq2}
\end{figure}

Figure \ref{fig:rq2} displays the results on four open-source websites.
The x-axis represents executed actions, and the y-axis represents visited web states.
The solid line indicates the mean and the shaded region represents the error bands.
Due to limited space, we only present the results for states here, as the trends of actions closely align with Figure \ref{fig:rq2}.
\tool consistently outperforms the other ablation groups.
Its superiority becomes more evident as the number of executed actions increased until a saturation is reached.

In Figure \ref{fig:rq2}, replacing MAB-based selector outperforms replacing input generator on DimeShift and Phoenix.
The latter requires successful login or registration for most functionalities, which demands higher input quality.
Even with a random selector, due to the limited interactive elements on the page, \tool can still complete registration thus achieves a promising result.
The same situation also applies to DimeShift, in which form filling is a predominant functional point.
We also conclude similar findings on the experiment on commercial websites.
Due to space limit, we do not further elaborate.

\begin{tcolorbox}
\textit{\textbf{Answer 2}}:
The LVLM-based components in \tool greatly enhance its web exploration capability.
\end{tcolorbox}

\subsection{RQ3: Tool Usefulness}

\subsubsection{Overview}

In the previous RQs, we already demonstrated \toolns's effectiveness by its exploration capabilities.
However, it does not sufficiently reflect \toolns's value in real-world software development.
Therefore, we conduct further investigations to assess the practical usefulness of \toolns.
They focus on two aspects: the quality of text inputs generated by \toolns's LVLM-based component, and the web failures and errors that were triggered during our experiment.

\subsubsection{Input Quality}

To assess the quality of input texts generated by \toolns, we recruit 5 graduate students with experience in GUI testing.
To ensure a 95\% confidence level and a 5\% margin of error, we randomly selected 329 samples from a total of 2,234 input texts collected during the experiment.

Each participant is provided with generated texts and the corresponding webpage screenshots with input box annotated, allowing them to visually observe the association between the text and the input box.
They are then asked to assess the input based on the input’s relevance to the element’s topic, using a 5-point Likert scale \cite{brown2011likert}.



The human assessment resulted in an average score of 4.27, indicating a positive overall evaluation.
Additionally, the Kappa coefficient \cite{wan2015kappa}, which is used to measure the inter-rater reliability among the participants, yielded a result of 0.63, implying a substantial agreement.
This demonstrates that \tool can generate high-quality text inputs for WUTs.

\subsubsection{Failures and Errors}

\begin{figure}[t!]
\centering
\includegraphics[width=\columnwidth]{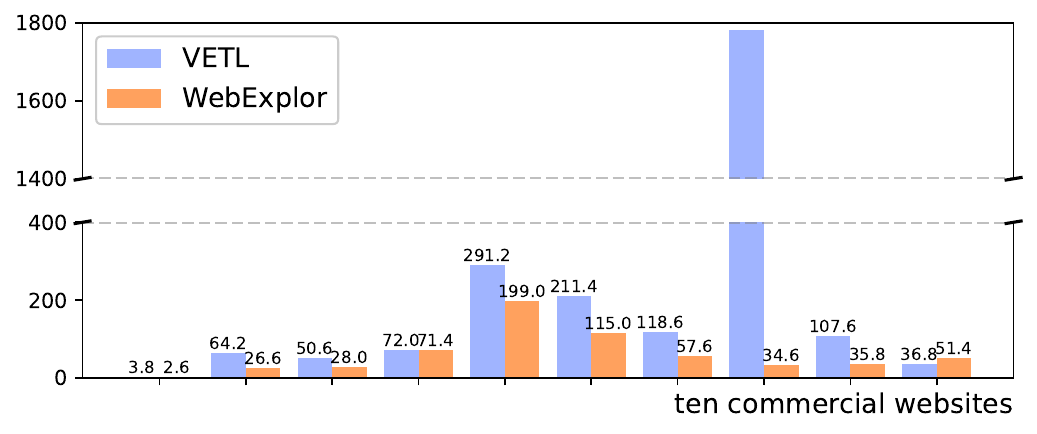}
\caption{Detected failures on commercial websites}
\label{fig:rq3-failure}
\end{figure}

We calculate the average numbers of failures triggered by \tool and WebExplor on commercial websites in our experiment.
They are collected from the failure logging messages on the web browser's console.
The result is shown in Figure \ref{fig:rq3-failure}.
In nine out of the ten commercial websites, \tool successfully triggers more failures than WebExplor, highlighting its effective failure detection capability.
Notably, in the website TvGuide \cite{tvguideGuideListings}, \tool triggers a huge number of failures, averaging at 1782.4.
We will revisit this case later.


Even though web failures can reflect the capability of an AWGT approach to trigger unknown web exceptions, the failures are not necessarily associated with severe bugs.
Here we present some real bugs that were identified upon our manual inspection, and also confirmed by the websites' maintainers.
In particular, we recorded videos for these identified bugs, which can be found in our public repository.

\begin{figure}[h!]
\centering
\includegraphics[width=0.99\columnwidth]{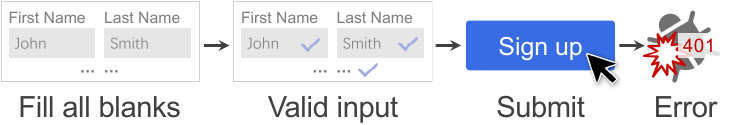}
\caption{An error (HTTP status code: 401) in Foursquare}
\label{fig:foursquare-bug}
\end{figure}

\noindent\textbf{$\bullet$ Failed user registration:}
Figure \ref{fig:foursquare-bug} illustrates a bug of broken business logic in Foursquare \cite{foursquareLocationTechnology}.
It causes user registration failures when the CAPTCHA is not loaded.
To trigger this error, a testing agent should complete the form with valid user information, and, most importantly, click the sign up button. 
Compared to WebExplor's single-step decision (i.e., the Markov property of reinforcement learning), \tool can support end-to-end functionality testing, facilitating the discovery of such errors.

\noindent\textbf{$\bullet$ Exhausting resources:}
Figure \ref{fig:rq3-failure} shows that \tool can trigger a huge number of failures in the 8th WUT, TvGuide.
We found that 85.4\% of them are browser errors of type \texttt{ERR\_INSUFFICIENT\_RESOURCES}, which indicates an inefficient memory usage of the WUT.
These failures continuously popped out after the testing agent entered an exceptionally long text into the search bar.
In our experiments, such text inputs can even crash the web browser.
WebExplor fails to produce such inputs as it can only generate strings whose lengths are within a fixed range.

\begin{tcolorbox}
\textit{\textbf{Answer 3}}:
\tool can generate high-quality and context-relevant text inputs for WUTs, which enable it to discover real bugs in top commercial websites.
\end{tcolorbox}

\section{Discussions}

\subsection{Future Work Directions}



Our experiment demonstrates the immense potential of leveraging LVLM in an AWGT task.
Nevertheless, we admit that the technical design of \tool is still in its early stage and can be improved in several aspects, which can be the focus of future work.
Here, we outline several potential directions:

$\bullet$~\textbf{Cooperating stateful exploration strategies}:
\tool selects target elements with MAB, a stateless decision-making algorithm.
This allows it to efficiently balance exploration and exploitation without considering web state abstraction.
In contrast, a stateful algorithm would consider the temporal correlation between consecutive actions \cite{zheng2021automatic}.
This can potentially bring more meaningful and effective test sequences.

$\bullet$~\textbf{Extracting input constraints with large models}:
In Section \ref{sssec:constrain}, we extract input constraints by analyzing specific DOM attributes (Table \ref{tab:constraint_pattern}).
With LVLM's image recognition capability and LLM's language understanding ability, we can derive more implicit constraints.
For instance, if the \texttt{id} attribute (or the screenshot) of an input box indicates its purpose for phone numbers, we can infer that it only allows numerical input with particular patterns.


$\bullet$~\textbf{Employing more powerful LVLMs}:
While \tool by-design is decoupled from a specific LVLM, we only experimented it with the smallest-scale LLaVA instance due to limited resources.
Considering that various LVLMs, such as larger LLaVA models or GPT-4V \cite{gpt4v}, have demonstrated consistently superior performances, it is possible that the effectiveness of VETL can be further boosted by incorporating more advanced LVLMs in place of LLaVA-7B.

\subsection{Threats To Validity}

The threat to internal validity comes from the uncertainty of output contents produced by VLM and LLM.
Language models are essentially probabilistic models which generate natural language texts in a non-deterministic manner.
As such, to reduce the randomness of test results, we run each experimental setting with five independent trials and present their average metrics during evaluation.

The threat to external validity concerns the representativeness of our subject websites.
To fairly compare \tool and its variants, as well as the state-of-the-art AWGT approach WebExplor, we selected four open-source websites used by many existing research work \cite{zheng2021automatic,biagiola2019diversity,fan2023comprehensive,Stocco2016APOGENAP}.
Further, to evaluate whether \tool can test real-world websites, we run experiment on ten top-ranking commercial websites.

\section{Related Work}

\subsection{Automatic Software GUI Testing}

Similar to \toolns, most GUI testing approaches generate tests by simulating human interactions with GUI elements to form action sequences.
Due to different targeting platforms, they vary in terms of state abstraction and action selection methods, as well as the underlying exploration strategies.

Crawljax \cite{mesbah2012crawling} is an early-stage AWGT tool. It adopts an on-the-fly model generation manner to systematically explore a WUT having dynamic content.
SUBWEB \cite{biagiola2017search} and DIG \cite{biagiola2019diversity} adopt model-based methods to achieve their testing objectives.
Eskonen \cite{eskonen2020automating} et al. propose to leverage the image recognition capability of deep neural network to accomplish state abstraction of complicated web contexts.
WebExplor \cite{zheng2021automatic} and QExplore \cite{sherin2023qexplore} utilize Q-learning, a classic reinforcement learning algorithm, to guide action selection in AWGT.
They differ in various aspects, including state abstraction, text input generation, and error recovery strategy. 

As the official Android GUI testing tool, Monkey \cite{developers2012monkey} can only generate simple events like random pixel clicking.
In order to smartly explore the apps, researchers have proposed testing methods based on intelligent algorithms.
For example, Sapienz \cite{mao2016sapienz} employs a multi-objective searching strategy to optimize test sequences on Android apps.
Ape \cite{gu2019practical}, as a model-based approach, adopts a runtime model refinement method to improve model precision.
Humanoid \cite{Li2019HumanoidAD} learns from human interactions to prioritize test input actions.
QDroid \cite{vuong2019semantic}, Q-testing \cite{pan2020reinforcement} and Fastbot \cite{cai2020fastbot,lv2022fastbot2} guide Android app exploration with reinforcement learning algorithms.

\subsection{LLM-driven Software Testing}

QTypist \cite{qtypist} is the first LLM-based text input generation method for Android GUI testing.
For a given input box, QTypist automatically extracts its global and local context, and infers its desired category of input data.
With the extracted data, it synthesizes prompts following a series of linguistic patterns.
The synthesized prompts guide a fine-tuned LLM to intelligently generate semantic text inputs.
Later, the authors of QTypist proposed InputBlaster \cite{inputblaster}, which aims at producing crash-triggering text inputs for mobile apps via LLM.
InputBlaster demonstrates example inputs to an LLM following an in-context learning scheme, and requests the LLM to write programs for producing unusual text strings.

Besides text input generation, researchers have also adopted LLM in other software testing scenarios, like end-to-end mobile app testing \cite{liu2023make} and program fuzzing \cite{deng2023large,deng2023large2}.
However, to the best of our knowledge, none of the existing research work, except \toolns, employs LVLM in software testing tasks.

Recently, researchers have also devoted efforts to develop LVLMs designated to GUI applications.
Examples of these LVLMs include CogVLM and its specialized version CogAgent \cite{wang2023cogvlm,Hong_2024_CVPR}.
There are also LVLM-based GUI agents, such as WebLinx \cite{lù2024weblinx} or SeeClick \cite{cheng2024seeclick}.
They can automatically perform actions on GUI applications based on user instructions like ``Create a task for a Career Fair on Google Calendar''.
Unlike these agents, \tool does not need user instructions.
It can autonomously explore the entire website, which differs from the task-oriented nature of WebLinx or SeeClick.

\section{Conclusion}

In this paper, we introduce \toolns, a novel LVLM-driven AWGT technique.
\tool effectively tackles the challenges of text input generation and target interactive element detection in a conventional AWGT scenario, both of which are solved by visual and textual prompt queries to the same LVLM.
Experimental results show that \tool surpasses the state-of-the-art approach, WebExplor, as well as its own variants, in exploring the state space and action space of WUTs.
Moreover, it successfully uncovers functional bugs in top-ranking commercial websites.
Our work represents the first attempt to employ pre-trained LVLM models in end-to-end web testing, highlighting the promising potential of this research direction.

\section*{Acknowledgment}

We would like to thank the ICSME 2024 reviewers for their helpful feedback on this paper.
This work is supported by the National Natural Science Foundation of China (Grant Nos. 61932021 and 62372219).

\bibliographystyle{IEEEtran}
\bibliography{references}

\end{document}